\documentclass[twocolumn,showpacs,preprintnumbers,amsmath,amssymb]{revtex4}

\usepackage{graphicx}
\usepackage{dcolumn}
\usepackage{bm}

\newcommand{\pa}{\parallel}

\newcommand{\cD}{{\cal D}}
\newcommand{\demi}{\frac{1}{2}}
\newcommand{\la}{\langle}
\newcommand{\ra}{\rangle}
\newcommand{\E}{{\bf E}}

\newcommand{\be}{\begin{equation}}
\newcommand{\ee}{\end{equation}}
\newcommand{\ba}{\begin{array}}
\newcommand{\ea}{\end{array}}
\newcommand{\ds} {\displaystyle}
\newcommand{\tr}{{\rm tr}}
\newcommand{\er}[1]{\hbox{(\ref{#1})}}

\newcommand{\ndop}{\rho_t}             
\newcommand{\undop}{\widetilde{\rho}_t}  
\newcommand{\rdop}{r_t}  
\newcommand{\ms}{y_t}              

\newcommand{\dmy} {\,\!}            

\begin{document}

\title{Pathwise Solution of a Class of Stochastic Master Equations}

\author{I.~Kurniawan}
\author{M.R.~James}%
\affiliation{%
Department of Engineering,\\
Australian National University,\\
Canberra, ACT 0200, Australia.\\
Indra.Kurniawan@anu.edu.au\\
Matthew.James@anu.edu.au
}%
\date{\today}

\begin{abstract}
In this paper we consider an alternative formulation of a class of
stochastic wave and master equations with scalar noise that are used in quantum
optics for modelling open systems and continuously monitored
systems. The reformulation is obtained by applying J.M.C.~Clark's
{\em pathwise} reformulation technique from the theory of
classical nonlinear filtering. The pathwise versions of the
stochastic wave and master equations are defined for all driving
paths and depend continuously on them.  In the case of white noise
equations, we derive analogs of Clark's {\em robust}
approximations. The results in this paper may be useful for
implementing filters for the continuous monitoring and measurement
feedback control of quantum systems, and for developing new types
of numerical methods for unravelling  master equations. The main
ideas are illustrated by an example.

\end{abstract}

\maketitle

\section{Introduction}
\label{sec:intro}

In quantum optics stochastic wave and master equations arise in
the study of open systems and continuous measurement, see, e.g.
\cite{GZ00}, \cite{HC93}, \cite{WM93}, \cite{BB91}, \cite{VPB92}, and the many references cited in these works.
These equations are driven by stochastic inputs, typically white
noise (Wiener process), representing photocurrent, or Poisson jumps, representing photon counts, and involve stochastic
integrals---they are stochastic differential equations (SDEs),
\cite{CWG04}, \cite{KS88}. These SDEs can be solved or approximated
numerically either for use in simulating open system dynamics, or
for updating conditional states (the topic of numerical
approximation of SDEs is well documented \cite{KP92},
\cite[Chapter 10]{CWG04}). However, it is important to keep in mind
that these are idealized models (e.g. the Wiener process is highly
irregular, in fact nowhere differentiable with probability one),
and the models may be used in conjunction with real data. Hence it
is of interest to consider  the robustness of stochastic wave and
master equations from a practical point of view.

Statistical robustness was  considered in 1978 by J.M.C.~Clark
\cite{JMCC78} in the context of classical nonlinear filtering.
 The theory of nonlinear filtering is an
important and well documented part of the systems and control,
communications, signal processing and probability and statistics
literature. It is well known (see, e.g. \cite{RE82, WH85}) that
the solution to the nonlinear filtering problem, say for a
diffusion state process observed in white noise, is given in terms
of the conditional distribution which solves a measure-valued
stochastic differential equation (analogous to the stochastic
mater equation). The corresponding equations for the conditional
density is a stochastic partial differential equation. Clark drew
attention to the disadvantages of stochastic integral
representations of nonlinear filters from a practical point of
view. These disadvantages concerned lack of statistical robustness
and the inability to cope with the range of measurement data
(driving process) that can arise in practice. Clark addressed
these issues by providing a reformulation of the nonlinear
filtering equations that does not involve stochastic integrals.
Clark's so-called {\em pathwise solution} defines a version of the
conditional distribution (or density) that is defined for all
possible measurement data and is a continuous function of the
measurement data, thereby providing important {\em robustness}
qualities. Clark also provided numerical approximations to the
reformulated nonlinear filters which inherit the robustness
characteristics. For further details, see \cite{JMCC78, MHAD79}, and related matters \cite{HJS78}.

In this paper we give reformulations of the stochastic wave and
master equations with scalar noise that do not involve stochastic integrals,
analogous to the classical pathwise versions of nonlinear filters
proposed by Clark. The reformulated equations are ordinary
differential equations where the driving process enters as a
parameter---not via a stochastic integral. This reformulation may
be useful for implementing filters for the continuous monitoring
and measurement feedback control of quantum systems, and for
developing new types of numerical methods for
\lq\lq{unravelling}\rq\rq \ master equations.

 The paper is organized as follows. In Section
\ref{sec:qfe} we describe the stochastic wave and master equations to be
considered, and provide some motivation and background
information. Then in Section \ref{sec:diff}, the pathwise solution
and robust approximation for quantum diffusion case are presented,
and we give an example to illustrate the solution in the context
of an imperfectly observed two-level atom continuously monitored
by homodyne photodetection. Section \ref{sec:jump} contains the
formulation for quantum jump case with some brief comments. Some calculations and the
proof of a continuity result are provided in  the Appendices. Some of
the results in this paper were announced in the conference paper
\cite{KJ04}.

\section{Stochastic Wave and Master Equations}
\label{sec:qfe}

\subsection{Background}
\label{sec:qfe-bg}

We recall (see, e.g. \cite[Chapter 2]{NC00}, \cite{EM98}) that an
isolated quantum system is described by a (pure) state $\vert \psi
\rangle \in \mathbf H$ (Dirac bra-ket notation), where $\mathbf H$
is a complex Hilbert space, with time evolution governed by the
Schrodinger equation
\be i \hbar \frac{\partial}{\partial t}\vert
\psi_t \rangle = H\vert \psi_t \rangle
\label{schrodinger}
\ee
where $\hbar=h/2\pi$ and $h$ is Planck's constant, and $H$ is a
Hamiltonian operator. In what follows we use units such that
$\hbar=1$.

However, when a quantum system is interacting with an external
environment, the interactions must be taken into account. In the
open systems literature (see, e.g. \cite[Chapter 5.4]{GZ00},
\cite[Chapter 6]{WM94}), the Schrodinger equation \er{schrodinger} is replaced
by a master equation, which takes the form
\be
\dot \rho = -i [H,\rho] + \cD[L]\rho ;
\label{master}
\ee
here  $\rho$ is the density operator and the
superoperator
$\cD$ is defined for any operator $c$ by
$$
\cD[c]\rho = c\rho c^\dagger - \demi c^\dagger c \rho - \demi \rho
c^\dagger c .
$$
The system operator $L$ is used in the modelling of the
interaction. Note that when $L=0$ and $\rho = \vert \psi \ra \la
\psi \vert$, the master equation \er{master} reduces to the
Schrodinger equation \er{schrodinger}.

A common method for solving the master equation \er{master} (via
\lq\lq{unravelling}\rq\rq) is to first solve a stochastic wave
equation and then to average. For instance, one could solve the
linear equation
\be
d \vert \tilde\psi \ra + K \vert \tilde\psi \ra dt = L \vert
\tilde\psi \ra dy
\label{stoch-wave}
\ee
for an unnormalized state $\vert \tilde\psi \ra$, where
\be
K = i H + \demi L^\dagger L ,
\label{K-def}
\ee
and $y(t)$ is a Wiener process, and then average
$$
\rho(t) = \E[ \vert \psi(t) \ra \la
\psi(t) \vert ] ,
$$
where
$$
\vert \psi(t) \ra = \frac{\vert \tilde\psi(t) \ra }{\sqrt{\la \tilde\psi(t) \vert \tilde\psi(t)
\ra}} \ .
$$
Such a procedure is computationally advantageous since the wave
function contains fewer components than the density operator.

Stochastic wave equations of the form \er{stoch-wave} and related
stochastic master equations also arise when quantum systems are
continuously monitored \cite[Chapter 11]{GZ00}, \cite{VPB92}. To
motivate this,  we recall that an ideal measurement of the system
is characterized by a self-adjoint operator $A$ on $\mathbf H$. In
the simple case that
$A$ has a discrete non-degenerate spectrum $\{ a_i \} \subset
\mathbf R$, the possible outcomes of a measurement are the
eigenvalues $a_i$. The outcome is {\em random}, where $a_i$ occurs
with probability
$$
p_i = \vert \langle a_i \vert \psi \rangle \vert^2
$$
when in state $\vert \psi \rangle $ (assumed normalized: $\vert
\psi \vert^2 = \tr[\vert \psi \rangle \langle \psi \vert]=1$ ).
Here, $\vert a_i \rangle$ denotes the orthonormal eigenvector of
$A$ corresponding to the eigenvalue $a_i$. After the measurement,
there is a collapse of the state to a new state \be \vert \psi_i'
\rangle = \vert a_i \rangle \langle a_i \vert \psi
\rangle/\sqrt{p_i} . \label{collapse} \ee The state $\vert \psi_i'
\rangle$ is the conditional state given the measurement outcome
$a_i$. Consequently, when a quantum system is measured, the
deterministic evolution given by the Schrodinger equation
\er{schrodinger} must be augmented by a stochastic transition,
e.g. \er{collapse}.  The continuous measurement of quantum systems
can be regarded in terms of a sequence of measurements of
infinitesimal strength (in contrast to the possibly large jump in
\er{collapse}) which accumulate in the limit to provide the
conditional information and the evolution of conditional states as
in \er{stoch-wave} and related stochastic master equations (see
below), \cite{CM87}, \cite[Chapter 11]{GZ00}.

In this paper we consider two kinds of stochastic master equation
corresponding to the two standard types of stochastic integrator
with independent increments: the standard Brownian motion
(diffusion case) and the standard Poisson type (jump case).

\subsection{Quantum Diffusion}
\label{sub:diff}

We consider a stochastic master equation (SME)
    \begin{eqnarray}
    d\ndop&=&[L\ndop L^\dag-K\ndop -\ndop K^\dag]dt\nonumber\\
    &&{+}\frac{1}{\kappa}\left[L\ndop +\ndop L^\dag-\ndop
    M_{\rho_t}\right]d\nu_t.
    \label{normd}
    \end{eqnarray}
In \er{normd}, the operator $K$ is defined by \er{K-def}. In the
case of continuous measurements, the parameter $\kappa \geq 1$ is
related to a measurement efficiency parameter $0 < \eta \leq 1$
via
$\kappa = 1/\sqrt{\eta}$ corresponding to imperfect or noisy
measurement; here perfect measurement corresponds to $\eta=1$.

The SME \er{normd} is driven by real valued white noise $\dot
\nu_t$, represented in \er{normd} by an Ito-sense stochastic
integral with respect to a standard Brownian motion (Wiener
process) $\nu_t$, sometimes called an innovations process. This
process is related to a real valued process $y_t$, which we call
the measurement process, by
   \be
    d\ms=M_{\rho_t}dt+\kappa d\nu_t,
    \label{meas}
    \ee
where $M_{\rho}{=}\langle
L+L^\dag\rangle_{\rho}=\text{tr}\{(L+L^\dag)\rho\}.$

If $\bar
\rho_t$ denotes the expected value of $\rho_t$, then $\bar \rho_t$
solves the  master equation \er{master}. Note that in the case
$L=0$ (no measurement or interaction) and initial pure state $\rho_0 = \vert
\psi_0 \rangle \langle \psi_0 \vert$, the state is pure for all
$t$, $\rho_t = \vert \psi_t \rangle \langle
\psi_t \vert$, and \er{normd} reduces to the
Schrodinger equation for $\vert \psi_t \rangle$ \er{schrodinger}.

The SME \er{normd} is nonlinear in $\rho_t$ (due to the term
$\rho_t M_{\rho_t}$), and the solution $\rho_t$ of \er{normd} is
normalized for all $t$: $\tr{\rho_t}=1$. We find it convenient to
work with an unnormalized version $\widetilde \rho_t$, defined by
$$
\widetilde \rho_t = \Lambda_t \rho_t
$$
where
   \be
    \Lambda_t=\exp\left\{\ds\frac{1}{\kappa^2}\left[\int_0^tM_{\rho_s}dy_s
    -\frac{1}{2}\int_0^t\bigl(M_{\rho_s}\bigr)^2ds\right]\right\}.
    \label{RN}
    \ee
It can be checked using Ito's rule (see, e.g. \cite[Chapters 12
and 18]{RE82}, \cite[Chapters 6 and 7]{WH85}) that $\widetilde
\rho_t$ solves the following linear stochastic equation: \be
    d\undop=[L\undop L^\dag-K\undop -\undop K^\dag]dt
    +\frac{1}{\kappa^2}\left[L\undop +\undop L^\dag
    \right]d\ms.
    \label{unormd}
    \ee
Note that this unnormalized SME is simpler and in bilinear
stochastic form driven by the measurement process $dy_t$. The
unnormalized density operator $\undop$  can be normalized by
simply dividing by its trace ($\Lambda_t=\tr(\widetilde \rho_t)$).
Equation \er{unormd} is analogous to the Duncan-Mortensen-Zakai
equation of nonlinear filtering (\cite[Chapter 18]{RE82},
\cite[Chapter 7]{WH85}), and is also known in the quantum physics
literature (\cite[Section 4]{GM96}, \cite[Section 4.2.2]{VPB01}).

Note that in the case
$\kappa=1$ and initial pure states, the unnormalized SME
\er{unormd} reduces to the unnormalized stochastic Schrodinger
equation \er{stoch-wave}; see \cite[Section 4.2.1]{VPB01},
\cite[Chapter 11]{GZ00}.

\subsection{Quantum Jumps}
\label{sub:jump}

Another type of continuous measurement is described by the
counting observation which give rise to a jump stochastic master
equation
    \begin{eqnarray}
    d\ndop&=&\bigl[-G\ndop-\ndop G^\dag+(1-\eta)\lambda\mathcal{J}\!\!\ndop
    +\eta\lambda\ndop\tr(\mathcal{J}\!\ndop)\bigr]dt\nonumber\\
    &&+\left[\frac{\mathcal{J}\!\ndop}{\tr(\mathcal{J}\!\ndop)}-\ndop\right]dN_t.
    \label{normd-j}
    \end{eqnarray}
The operator $G$ is defined by
    \be
    G=\frac{\lambda}{2}C^\dag C + iE
    \ee
where $C$ is some Schrodinger picture system operator, $E$ is
energy operator related to system Hamiltonian by
$H=E+\frac{i\lambda}{2}(C-C^\dag)$, $\lambda>0$ is related to the
intensity of the standard Poisson process and
$\mathcal{J}\!\ndop=C\ndop C^\dag$ is defined as the jump
superoperator. The SME \er{normd-j} is driven by the counting
observation process $dN_t$ which is a real random variable
satisfying
    \be\ba{rl}
    E[dN_t]&=\eta\lambda\,\tr(\mathcal{J}\!\ndop)\,dt\\
    dN_t^2 &= dN_t
    \ea\label{dNt}
    \ee
The observation process $dN_t$ only gives the value of either zero
or one (corresponding to the counting increment), and is the
representation of the standard Poisson process with intensity
$\eta\lambda\tr(\mathcal{J}\!\ndop)\,dt$. Here we assume that the
observation process has efficiency $0<\eta\leq 1$ as the
representation of imperfect or erroneous counting process.

The SME \er{normd-j} together with \er{dNt} simply tell us the
behavior of quantum jump that in the increment of time $dt$ the
system jumps via superoperator $\mathcal{J}$ with probability
$P_j=E[dN_t]$ or smoothly evolves via the first bracket term of
RHS of \er{normd-j} with probability $P_s = 1-P_j$. Note that the
average or expectation value of $\ndop$ in \er{normd-j} obeys
master equation \er{master} by considering the operator
$L=\lambda^{1/2}(C-I)$.

One can check for perfect counting process $(\eta=1)$ and initial
pure states that the jump SME \er{normd-j} reduces to the
normalized \emph{jump stochastic Schrodinger equation} (see e.g.
\cite{VPB01,BB91,WM93a})
    \be\ba{rl}
    d\vert\psi_t\rangle=&\ds\left[\frac{\lambda}{2}(\vert C\psi_t\vert^2-C^\dag
    C)-iE\right]\vert\psi_t\rangle\,dt\\
    &+\left[\ds\frac{C}{\vert C\psi_t\vert}-I\right]\vert\psi_t\rangle\,dN_t
    \ea\label{sse-j}\ee

The jump SME \er{normd-j} and \er{sse-j} are normalized but
nonlinear so again we work with unnormalized version $\widetilde
\rho_t$, defined by $$\widetilde \rho_t = \Lambda_t \rho_t$$ where
    \be
    \Lambda_t=1 + \ds\int_0^t\Delta_s\tr(\mathcal{J}\!\rho_s-\rho_s)\left[dN_s
    -\eta\lambda\,ds\right].
    \label{RN-j}
    \ee
And by Ito's rule for jump process, the unnormalized $\undop$
solves the following linear stochastic jump equation
    \be\ba{rl}
    d\undop=&[-G\undop -\undop
    G^\dag+(1-\eta)\lambda\mathcal{J}\!\undop+\eta\lambda\undop]dt\\
    &+\left[\mathcal{J}\!\undop-\undop\right]dN_t.
    \label{unormd-j}
    \ea\ee

We see that the unnormalized jump SME \er{unormd-j} reduces to an
unnormalized and linear jump stochastic Schrodinger equation
    \be\ba{rl}
    d\vert\widetilde{\psi}_t\rangle=\left[\frac{\lambda}{2}(I-C^\dag
    C)-iE\right]\vert\widetilde{\psi}_t\rangle\,dt
    +\left[C-I\right]\vert\widetilde{\psi}_t\rangle\,dN_t
    \ea\label{lin-sse-j}\ee
for the case of perfect measurement and pure initial states.

\section{Diffusion Case}
\label{sec:diff}

\subsection{Pathwise Solution}
\label{sub:pathwise}

We follow Clark's approach \cite{JMCC78} to obtain a pathwise
solution to the SME \er{normd}. Let \be
    A_t=\exp\left\{\!-\frac{L}{\kappa^2}\ms+\frac{L^2}{2\kappa^2}t\right\} .
    \label{A}
    \ee
Let $\widetilde \rho_t$ be a solution of the unnormalized
stochastic master equation \er{unormd}, and define an unnormalized state $r_t$ by
\be
    \rdop=A_t\undop A^\dag_t .
    \label{transf}
\ee
Then, as shown in Appendix \ref{app:robust},  $r_t$ solves the
{\em pathwise master equation}
    \be
    \dot{\rdop}=L\rdop L^\dag\!\left[1-1/\kappa^2\right]
    -A_tKA^{-1}_t\rdop-\rdop (A^\dag_t)^{-1}K^\dag A^\dag_t .
    \label{rb}
    \ee
Conversely, solutions $\rho_t$, $\widetilde \rho_t$ to the
stochastic master equations \er{normd}, \er{unormd} can be
obtained from a solution $r_t$ to \er{rb} via the formulas
\be
\widetilde \rho_t = A^{-1}_t r_t (A^\dagger_t)^{-1}, \ \ \rho_t =
\frac{A^{-1}_t r_t (A^\dagger_t)^{-1}}{\tr[A^{-1}_t r_t
(A^\dagger_t)^{-1}]} .
\label{robust-def}
\ee

This result is analogous to the classical result for nonlinear
filtering \cite[Theorems 4 and 6]{JMCC78}. It is important to note
that the pathwise equation \er{rb} does not involve stochastic
integrals; the measurement path enters as a parameter in an
ordinary equation. In particular, the versions of the solutions to
the stochastic master equations \er{normd}, \er{unormd} defined by
\er{rb}, \er{robust-def} are defined for all continuous
observations paths, not just for a set of paths of full Wiener
measure as is the case for solutions obtained directly due to the
stochastic integrals in \er{normd}, \er{unormd}.

We next  make explicit the continuous dependence on the
measurement paths. We make use of the supremum norm
$$
\pa f \pa_T = \sup_{0 \leq t \leq T} \vert f_t \vert
$$
for a continuous vector or matrix valued function, and $\vert
\cdot \vert$ denotes the appropriate Euclidean norm.

Let ${\bf H}$ be finite dimensional. Then, as shown in Appendix
\ref{app:proof},  solutions $\widetilde\rho_t$, $\rho_t$ to the
stochastic master equations \er{unormd}, \er{normd} defined by
\er{robust-def} are locally Lipschitz continuous functions of the
observation trajectories. This means that if $y^1_t$ and $y^2_t$
are two continuous observation trajectories on $[0,T]$, with
corresponding solutions $\widetilde \rho^1_t$,  $\rho^1_t$and
$\widetilde \rho^2_t$, $\rho^2_t$, respectively, then there exists
a positive constant $C=C(\pa y^1\pa_T,\pa y^2\pa_T,T)$ such that
\begin{eqnarray}
\pa \widetilde\rho^1 - \widetilde\rho^2 \pa_T &\leq & C \pa y^1 - y^2 \pa_T , \ \text{and} \
\nonumber \\
\pa \rho^1 - \rho^2 \pa_T &\leq & C \pa y^1 - y^2 \pa_T .
\label{rho-cts}
\end{eqnarray}

We note that in the case of perfect measurement and initial pure
states, the pathwise SME \er{rb} reduces to a  {\em pathwise
Schrodinger equation}:
\be \dot{\vert \phi_t \rangle} = - A_t K
A_t^{-1}\vert \phi_t \rangle .
\label{rb-schrodinger}
\ee
This equation is also defined for all continuous observation
paths, and depends continuously on them.

\subsection{Robust Approximation}
\label{sub:robust}

In this section we use the pathwise master equation \er{rb} to
derive an approximation to the stochastic master equation
\er{normd}. We will restrict our attention to the case of a finite
dimensional underlying Hilbert space, and we employ a simple
implicit Euler scheme to illustrate the ideas. In \cite{JMCC78},
the corresponding approximations for nonlinear filters were called
{\em robust approximations}.

Fix the interval between sampling times $\Delta{=}t_n-t_{n-1}$.
A reasonable implicit Euler approximation for \er{rb} is
    \begin{eqnarray}
    \rdop\dmy^\Delta_n&=&\rdop\dmy^\Delta_{n-1}+L\rdop\dmy^\Delta_n
    L^\dag\left[1-1/\kappa^2\right]\Delta\nonumber\\
    & &-\:[A_{t_n}KA_{t_n}^{-1}\rdop\dmy^\Delta_n+\rdop\dmy^\Delta_n (A_{t_n}^\dag)^{-1}K^\dag
    A^\dag_{t_n}]\Delta.
    \nonumber
    \label{discr}\end{eqnarray}
Multiplying both sides by
$A_{t_n}^{-1}\{\ldots\}(A_{t_n}^\dag)^{-1}$ gives
    \begin{eqnarray}
      \undop\dmy^\Delta_n&=&A_{t_n}^{-1}A_{t_{n-1}}  \undop\dmy^\Delta_{n-1}A_{t_{n-1}}^\dag
    (A_{t_n}^\dag)^{-1}\nonumber\\
    & &+\:L  \undop\dmy^\Delta_n
    L^\dag\left[1-1/\kappa^2\right]\Delta\nonumber
    -[K  \undop\dmy^\Delta_n+  \undop\dmy^\Delta_n K^\dag]\Delta.
    \nonumber
    \end{eqnarray}
Defining $\Delta y_n=y_{t_n}-y_{t_{n-1}}$, we write
    \begin{eqnarray}
    A_{t_n}^{-1}A_{t_{n-1}}&=&\exp{\left\{\frac{L}{\kappa^2}\Delta
    y_n-\frac{L^2}{2\kappa^2}\Delta\right\}} = \mathcal{E}(\Delta y_n)\nonumber\\
    A^\dag_{t_{n-1}}(A^\dag_{t_n})^{-1}&=&\exp{\left\{\frac{L^\dag}{\kappa^2}\Delta
    y_n-\frac{(L^\dag)^2}{2\kappa^2}\Delta\right\}}=\mathcal{E}(\Delta y_n)^\dagger.\nonumber
    \end{eqnarray}
We obtain the implicit robust approximation for the unnormalized
SME expressed as a matrix equation
    \be
    \mathcal{A}  \undop\dmy^\Delta_n+  \undop\dmy^\Delta_n\mathcal{B}-\mathcal{C}  \undop\dmy^\Delta_n\mathcal{D}
    =\mathcal{E}(\Delta y_n)  \undop\dmy^\Delta_{n-1}\mathcal{E}(\Delta y_n)^\dagger,
    \label{mtx}
    \ee
where
    \begin{eqnarray}
    \mathcal{A}&{}={}&[I+K\Delta]\nonumber\\
    \mathcal{B}&{}={}&K^\dag\Delta\nonumber\\
    \mathcal{C}&{}={}&L\nonumber\\
    \mathcal{D}&{}={}&L^\dag\left[1-1/\kappa^2\right]\Delta .\nonumber
    \end{eqnarray}
The matrix equation \er{mtx} can be solved explicitly by
rearranging elements of the matrix. Suppose
$\mathcal{A}=[A_1|A_2|\ldots|A_n]$ is an $n{\times}n$ matrix,
$A_1\ldots A_n$ are column vectors of $\mathcal{A}$. We define
operator $\mathrm{Vec}(\mathcal{A})=[A_1|A_2|\ldots|A_n]^T$. Thus,
Vec operator transforms $n{\times}n$ matrix to $(nn){\times}1$
matrix.

Suppose $\mathcal{B}$ and $\mathcal{X}$ are also $n{\times}n$
matrices, then
    \be
    \mathrm{Vec}[\mathcal{AXB}]=[\mathcal{B}^T\otimes
    \mathcal{A}]\mathrm{Vec}\mathcal{X},
    \ee
where $\otimes$ denotes Kronecker product. If $\mathcal{B}$ is a
complex matrix, then $\mathcal{B}^T$ is simply transposing
$\mathcal{B}$ \emph{without} conjugating it.

Transform \er{mtx} by Vec operator
    \begin{eqnarray}
    \mathrm{Vec}(\mathcal{A}  \undop\dmy^\Delta_n)+\mathrm{Vec}(  \undop\dmy^\Delta_n\mathcal{B})
    -\mathrm{Vec}(\mathcal{C}  \undop\dmy^\Delta_n\mathcal{D}) =
\\
    \mathrm{Vec}(\mathcal{E}(\Delta y_n) \undop\dmy^\Delta_{n-1} \mathcal{E}(\Delta
    y_n)^\dagger)
    \nonumber\\
    \bigl[(I\otimes
    \mathcal{A})+(\mathcal{B}^T\otimes I)
    -(\mathcal{D}^T\otimes \mathcal{C})\bigr]
    \mathrm{Vec}(  \undop\dmy^\Delta_n) =
\\ \mathrm{Vec}(\mathcal{E}(\Delta y_n) \undop\dmy^\Delta_{n-1} \mathcal{E}(\Delta
    y_n)^\dagger),\nonumber
    \end{eqnarray}
yields
    \be\ba{rl}
    \mathrm{Vec}(  \undop\dmy^\Delta_n)=&
    \bigl[(I\otimes \mathcal{A})+(\mathcal{B}^T\otimes I)
    -(\mathcal{D}^T\otimes
    \mathcal{C})\bigr]^{-1}
    \\&\times\mathrm{Vec}(\mathcal{E}(\Delta y_n) \undop\dmy^\Delta_{n-1} \mathcal{E}(\Delta
    y_n)^\dagger) .
    \ea\label{impVec}
    \ee
Writing \er{impVec} in the symbolic form
    \be
      \undop\dmy^\Delta_n=\Gamma(\Delta y_n)  \undop\dmy^\Delta_{n-1},
    \label{rec}\ee
we obtain an approximation to the solution of the unnormalized
stochastic master equation \er{unormd}. By normalization we have
    \be
    \ndop\dmy^\Delta_n= \frac{\Gamma(\Delta y_n)  \ndop\dmy^\Delta_{n-1}}{\tr[ \Gamma(\Delta y_n)  \ndop\dmy^\Delta_{n-1}
    ]},
    \label{approx-sme}
    \ee
an approximation to the stochastic master equation \er{normd}.

Note that $\Gamma$ in \er{rec} can be seen as common recursive
filtering solution incorporating two steps, \emph{prediction }and
\emph{update} or \emph{correction}. The prediction step utilizes
knowledge in $\ms$ histories $(y_{t_{0:n-1}})$ via $
\undop\dmy^\Delta_{n-1}$ and then the result is updated by the
current measurement information available in $\Delta y_n$.

A significant result concerning robust approximations
\cite[Theorem 7]{JMCC78} is that the convergence of the
approximation  $\undop\dmy^\Delta_n$ to the exact solution $
\widetilde{\rho}_{t_n}$ is pathwise for all observation
trajectories. Indeed, the following inequality can be proven as in
\cite[Theorem 7]{JMCC78}: there exists a continuous function
$k(\cdot)$ such that for all $\Delta >0$ and $n$ with $0 \leq t_n
\leq T$, \be \vert \undop\dmy^\Delta_n(y) -
\widetilde{\rho}_{t_n}(y) \vert \leq k( \pa y \pa_T )(\Delta +
w_y(\Delta)) \label{delta} \ee where
$$
w_y(\Delta) = \max\{ \vert y(s_1)-y(s_2) \vert
\ : \ 0 \leq s_1, s_2 \leq \Delta     \}   .
$$
This should be compared with other discrete approximations to solutions for SDEs,  \cite{KP92},
\cite[Chapter 10]{CWG04}; for example, as discussed in \cite[Section 4]{JMCC78},  a direct Euler approximation of \er{unormd} converges \lq\lq{almost surely}\rq\rq, but for differentiable observation trajectories $y$, the Euler approximation  converges to a limit which is not the same as $\widetilde \rho_t(y)$ given by \er{robust-def}.

\subsection{Example}
\label{sub:eg1}

In this section, we apply the discrete approximation derived in
Subsection \ref{sub:robust} to a two level atom continuously
monitored by homodyne photodetection \cite[Section III.C]{WM93b}.
In this example, the underlying Hilbert space is $\mathbf C^2$,
the two-dimensional complex vector space, whose elements are
called {\em qbits} in quantum computing. Let $\vert 0 \rangle$ and
$\vert 1 \rangle$ denote basis vectors corresponding to ground and
excited states, respectively. We use the following Pauli matrices
to represent operators for this system:
 \setlength{\arraycolsep}{0.0em}
    \begin{eqnarray}
    \sigma_x&{}={}&|0\rangle \langle 1|+|1\rangle \langle 0|
    =\left[%
    \begin{array}{lr}
    0 \ \  & 1 \\
    1 & 0 \\
    \end{array}%
    \right]\nonumber\\
    \sigma_y&{}={}&i|0\rangle \langle 1|-i|1\rangle \langle 0|
    =\left[%
    \begin{array}{lr}
    0 \ \ & -i \\
    i & 0 \\
    \end{array}%
    \right]\nonumber\\
    \sigma_z&{}={}&|1\rangle \langle 1|-|0\rangle \langle 0|
    =\left[%
    \begin{array}{lr}
    1 \ \ & 0 \\
    0 & -1 \\
    \end{array}%
    \right]\nonumber\\
    \sigma &{}={}&|0\rangle \langle 1|=\frac{1}{2}(\sigma_x-i\sigma_y)
    =\left[%
    \begin{array}{lr}
    0 \ \ & 0 \\
    1 & 0 \\
    \end{array}%
    \right] .
    \label{pauli}
    \end{eqnarray}
    \setlength{\arraycolsep}{5pt}Here $\sigma$ is a system (lowering) operator. Any state $\rho$ on
$\mathbf C^2$ can be represented in terms of the Bloch vector
$(x,y,z)$ (\cite[Chapter 2]{NC00}):
\setlength{\arraycolsep}{0.0em}
    \begin{eqnarray}
    \rho&{}={}&\frac{1}{2}\left[I+x\sigma_x+y\sigma_y+z\sigma_y\right]\nonumber\\
    &{}={}&\frac{1}{2}
    \left[%
    \begin{array}{cc}
    1{+}z  &\  x{-}i y \\
    x{+}i y  &\  1{-}z \\
    \end{array}%
    \right],
    \label{Bloch}\end{eqnarray}
    \setlength{\arraycolsep}{5pt}where $x^2+y^2+z^2 \leq 1$. The Hamiltonian of the system is given
by \be H = \frac{\alpha}{2}\sigma_x+\frac{\Delta}{2}\sigma_z
    =\frac{1}{2}\left[%
            \begin{array}{cc}
            \Delta\  & \alpha \\
            \alpha\  & -\Delta \\
            \end{array}%
            \right]
\label{qbit-H} \ee where $\alpha$ is the Rabi frequency, $\Delta$
is the atomic frequency minus the classical field frequency.

The two level atom is coupled to an optical field which is
continuously monitored by homodyne detection (see, e.g.
\cite[Section 8.7]{HB98}, \cite[Section II.C]{WM93a}). The output
of the detector is a current $I_c(t)$ whose mean value is
proportional to the expected field quadrature $\tr[ X_\varphi \rho
]$ determined by the phase angle $\varphi$ of the local
oscillator. Here,
$$
X_\varphi = \frac{1}{2} ( e^{-i\varphi} \sigma + e^{i\varphi}
\sigma^\dagger ).
$$
For $\varphi=0$ we are interested in measuring the $x$-quadrature
$2X_0 = \sigma_x$, while for $\varphi=\pi/2$ we are interested in
measuring the $y$-quadrature $2X_{\pi/2} = -\sigma_y$. Variations
about the mean are called {\em quantum noise}, a key feature of
quantum optical systems.

The stochastic master equation for this setup is equation
\er{normd} with $H$ given by \er{qbit-H} and \be L= \sqrt{\gamma}
e^{-i\varphi} \sigma , \label{qbit-L} \ee where $\gamma$ is the
spontaneous emission rate. In this case
$K=\frac{\gamma}{2}\sigma^\dag\sigma+iH$. The corresponding
measurement equation is \er{meas}, where $I_c(t) = \dot y_t$ is
the homodyne photocurrent. The quantum noise $(1/\sqrt{\eta}) \dot
\nu_t$ is white with variance $1/\eta$.

The approximation \er{approx-sme} was implemented for this example
with an assumed detection efficiency of $\eta=85\%$ (a low value),
to compare with the results of \cite[section III.C]{WM93a} which
used different methods and considered the case of perfect
measurement efficiency $\eta=1$. The simulation was carried out as
follows:
\begin{itemize}
    \item Set $\gamma{=}1$. All other parameters are based on
    $\gamma$ unit, $\Delta{=}0$, $\alpha{=}\frac{7}{\sqrt{2}}\gamma$.
    Simulations are conducted with two values of $\varphi{=}0$ and $\varphi{=}\pi/2$.
    \item Time step $\Delta{=}0.01/\gamma$, time length
    $T{=}25/\gamma$ correspond to simulation length $n{=}2500$.
    \item Simulations are done for single ensemble and $N{=}1000$ ensembles.
    \item Set pure state initial condition of $|\psi_0\rangle{=}\frac{|0\rangle+|1\rangle}{\sqrt{2}}$
    such that
    $\rho_0{=}|\psi_0\rangle\langle\psi_0|{=}\frac{1}{2}\left[%
                    \begin{array}{cc}
                    1 & 1 \\
                    1 & 1 \\
                    \end{array}%
                    \right]$. This corresponds to initial Bloch
                    vector $(x,y,z)_0=(1,0,0)$.
    \item We compute recursively $\ndop$
    via \er{approx-sme}. The new measurement data generated by
    $$\Delta y_{n}=\text{tr}\{(L+L^\dag)\ndop\dmy^\Delta_{n-1}\}\Delta+\kappa
    \Delta\nu_n ,
    $$
    where $\Delta\nu_n$ is an independent identically distributed
    Gaussian sequence with mean zero and variance $\Delta$.
    \item We obtain the
    corresponding Bloch vector  $(x_{tn},y_{tn},z_{tn})$ by using \er{Bloch}.
\end{itemize}

\begin{figure}
\centering
\includegraphics[width=2.7in]{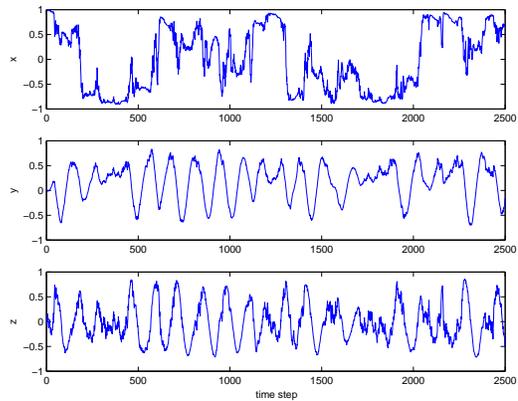}
\caption{The evolution of one ensemble Bloch vector with
$\varphi{=}0$} \label{fig1}
\end{figure}

In spite of the poor measurement efficiency, one can still infer
important physical information about the system as described  in
\cite[section III.C]{WM93b}. Indeed, in terms of the Bloch vector
$(x_t,y_t,z_t)$, the homodyne photocurrent is
$$
I_c(t)=\sqrt{\gamma}[x_t\cos\varphi-y_t\sin\varphi]+\kappa \dot
\nu_t.
$$
When the local oscillator is in phase with the driving field
$(\varphi{=}0)$, the deterministic part of measurement is
proportional to $\langle\sigma_x\rangle$. This measurement seems
to drive the system into an eigenstate of $\sigma_x$. This is
shown in Fig.\ref{fig1}. Alternatively, one can see from the
steady state ensembles simulation Fig.\ref{fig3} that the atom
states are concentrated near $x=+1$ and $x=-1$.

\begin{figure}
\centering
\includegraphics[width=2.7in]{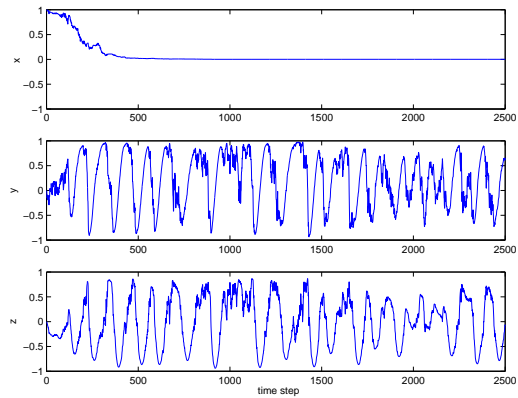}
\caption{The evolution of one ensemble Bloch vector with
$\varphi{=}\pi/2$} \label{fig2}
\end{figure}

In contrast, measuring the quadrature with $\varphi=\pi/2$ will
eventually force the atom states into the eigenstates of
$\sigma_y$. The states are spinning around the sphere toward
$z=+1$ and $z=-1$ due to the driving Hamiltonian, Fig.\ref{fig2},
Fig.\ref{fig4}. 

The effect of imperfect measurement can be seen clearly from these
results. The Bloch vectors are not  confined to the surface of the
unit sphere, thus the system is in mixed states. This shows that
the imperfect measurements can cause loss of information about the
quantum system, but nevertheless the information is consistent
with the perfect case \cite[section III.C]{WM93b}.

\begin{figure}
\centering
\includegraphics[width=2.7in]{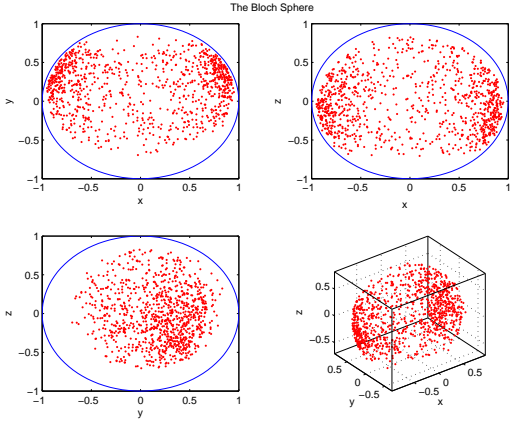}
\caption{The steady state point of 1000 ensembles Bloch vector on
the Bloch sphere with $\varphi{=}0$} \label{fig3}
\end{figure}

\begin{figure}
\centering
\includegraphics[width=2.7in]{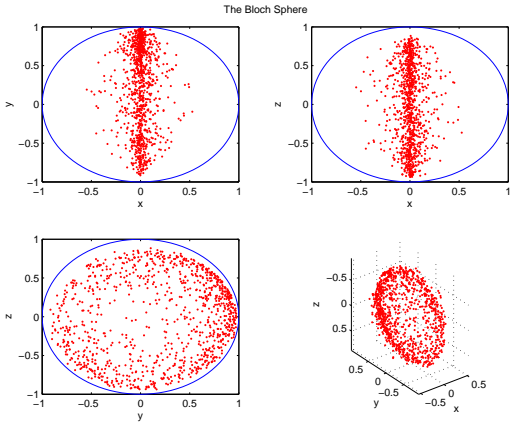}
\caption{The steady state point of 1000 ensembles Bloch vector on
the Bloch sphere with $\varphi{=}\pi/2$} \label{fig4}
\end{figure}

\section{Jump Case}
\label{sec:jump}

To derive a pathwise solution for the jump SME \er{normd-j}, we
choose (see for example \cite{MEJ99})
    \be
    A_t=C^{-N_t},
    \label{A-j}
    \ee
and again define an unnormalized state $r_t$ by
    \be
    \rdop=A_t\undop A^\dag_t ,
    \label{transf-j}
    \ee
where $\undop$ is a solution of the unnormalized jump stochastic master equation
\er{unormd-j}. Then $r_t$ solves the {\em pathwise master
equation}
    \be
    \dot{\rdop}=-A_tGA^{-1}_t\rdop-\rdop (A^\dag_t)^{-1}G^\dag A^\dag_t
    + (1-\eta)\lambda\mathcal{J}\!\rdop + \eta\lambda\rdop.
    \label{rb-j}
    \ee
Moreover, solutions $\rho_t$, $\widetilde \rho_t$ to the
stochastic master equations \er{normd-j}, \er{unormd-j} can be
obtained from a solution $r_t$ to \er{rb-j} via the formulas
\be
\widetilde \rho_t = A^{-1}_t r_t (A^\dagger_t)^{-1}, \ \ \rho_t =
\frac{A^{-1}_t r_t (A^\dagger_t)^{-1}}{\tr[A^{-1}_t r_t
(A^\dagger_t)^{-1}]}.
\label{robust-def-j}
\ee

In the case of perfect measurement and initial pure states, the
jump SME \er{rb-j} admits the \emph{jump pathwise Schrodinger}
equation
\be \dot{\vert \phi_t \rangle} = \left(- A_t G
A_t^{-1}+\frac{\lambda}{2}I\right)\vert \phi_t \rangle .
\label{rb-schrodinger-j}
\ee

The pathwise solutions \er{rb-j} and \er{rb-schrodinger-j} appear
as ordinary equation without stochastic integrals in terms of
$dN_t$; the counting observation result $N_t$ enters as a
parameter in $A_t$. These solutions are also defined for all
continuous counting observation paths.

However, we note that for the case of the jump SME, the pathwise
solution might not be so useful for computation, and existing
techniques (see, e.g. \cite[Chapters 11]{GZ00}) may be preferable.

\section{Conclusion}
\label{sec:sum}

In this paper we have proposed the pathwise reformulation of the
stochastic master equations for two cases; quantum diffusion and
jump. These reformulations provide solution that is defined for
all measurement paths and enjoy continuity properties. These
robustness characteristic would be useful when applied to quantum
filtering problem such as in quantum feedback control.

The results we have established are valid for scalar measurements,
but can easily be generalized to the case of multiple measurements
provided  the interaction operators commute. It is not known if
the results generalize in case the interaction operators do not
commute.

\begin{acknowledgments}
We wish to acknowledge the support of AusAID and ARC for this
research.
\end{acknowledgments}

\appendix

\section{Proof of Pathwise Equations}
\label{app:robust}

In this appendix we prove the assertions of sections \ref{sub:pathwise} and \ref{sec:jump} concerning the pathwise equations.

We first consider the diffusion case and verify \er{rb} and \er{robust-def}.
 The calculations are  simple but needs frequent and careful use of
Ito's rule. Differentiate \er{transf},
    \begin{eqnarray}
    d\rdop&=&dA_t.\undop A_t^\dag + A_td\undop.A_t^\dag + A_t\undop.dA_t^\dag+A_td\undop.dA_t^\dag\nonumber\\
    &&+\:dA_t.\undop.dA_t^\dag + dA_t.d\undop.A_t^\dag +
    dA_t.d\undop.dA_t^\dag,
    \label{dr}\end{eqnarray}
and \er{A},
    \begin{eqnarray}
    dA_t&=&\ds\frac{\partial A_t}{\partial\ms}d\ms
    +\frac{\partial A_t}{\partial t}dt
    +\frac{1}{2}\frac{\partial^2A_t}{\partial\ms^2}\,d\ms^2\nonumber\\
    &=&
    -A_t\frac{L}{\kappa^2}d\ms+A_t\frac{L^2}{2\kappa^2}dt+\frac{1}{2}(\frac{-L}{\kappa^2})
    (-A_t\frac{L}{\kappa^2})\kappa^2dt\nonumber\\
    &=&-A_t\frac{L}{\kappa^2}d\ms+ A_t\frac{L^2}{\kappa^2}dt.
    \label{p1}\end{eqnarray}
Similarly,
    \be
    dA_t^\dag=-A_t^\dag\frac{L^\dag}{\kappa^2}d\ms+
    A_t^\dag\frac{(L^\dag)^2}{\kappa^2}dt.
    \label{p2}\ee
Put \er{unormd}, \er{p1}, \er{p2} together into \er{dr}, and
carefully using Ito's rule, we obtain \er{rb}.

In the jump case we prove \er{rb-j} and \er{robust-def-j} analogously to the diffusion case. We need to use Ito's rule for $dN_t$, i.e. $dN_t^2=dN_t$ and $dN_tdt=0$. Any
higher order of differential involving $dt$ is zero. Differentiate
\er{transf-j},
    \begin{eqnarray}
    d\rdop&=&dA_t.\undop A_t^\dag + A_td\undop.A_t^\dag + A_t\undop.dA_t^\dag+A_td\undop.dA_t^\dag\nonumber\\
    &&+dA_t.\undop.dA_t^\dag + dA_t.d\undop.A_t^\dag +
    dA_t.d\undop.dA_t^\dag,
    \label{dr-j}\end{eqnarray}
and \er{A},
    \begin{eqnarray}
    dA_t&=&A_t\left(\ds\sum_{n=1}^\infty(-1)^n\frac{(\ln
    C)^n}{n!}\right)\,dN_t\nonumber\\
    &=&A_t(C^{-1}-I)\,dN_t.
    \label{p1-j}\end{eqnarray}
The infinite series arise due to the Ito's rule that higher orders
of $dN_t$ do not vanish. Similarly,
    \be
    dA_t^\dag= A_t^\dag((C^\dag)^{-1}-I)\,dN_t
    \label{p2-j}
    \ee
Put \er{unormd-j}, \er{p1-j}, \er{p2-j} together into \er{dr-j},
and carefully using Ito's rule, we obtain \er{rb-j}.

\section{Continuity Proof}
\label{app:proof}

We provide the  proof of the continuity result \er{rho-cts} of section \ref{sub:pathwise}.

In view of \er{robust-def}, it is enough to verify that
\be
\pa r^1 - r^2 \pa_T \leq C \pa y^1 - y^2 \pa_T .
\label{cts}
\ee

 Let Equation
\er{rb} defined for all $t\in[0,T]$, and we rewrite it in term of
matrices $\mathcal{R}_t=\mathcal{R}(y_t)$ and $\mathcal{M}$ such
that
    \be
    \dot{\rdop}=\mathcal{R}_t\rdop+\rdop \mathcal{R}^\dag_t+
    \mathcal{M}(\rdop).
    \label{rb-p}
    \ee
 Given observation records $y_t^1$, $y_t^2$ we have the corresponding pathwise solutions
$r_t^1=r_t(y_t^1)$, $r_t^2=r_t(y_t^2)$ and matrices
$\mathcal{R}_t^1=\mathcal{R}(y_t^1)$,
$\mathcal{R}_t^2=\mathcal{R}(y_t^2)$.

It follows
    \begin{eqnarray*}
      \vert r_t^1-r_t^2\vert &\leq& \ds\int_0^t\left|\mathcal{R}_s^1r_s^1-
      \mathcal{R}_s^2 r_s^2 + r_s^1 \mathcal{R}_s^{1\dag}-r_s^2 \mathcal{R}_s^{2\dag}\right.\\
      && +\left.\mathcal{M}(r_s^1)-\mathcal{M}(r_s^2)\right|ds\\
       &\leq& \ds\int_0^t\left|\mathcal{R}_s^1r_s^1- \mathcal{R}_s^2r_s^1+ \mathcal{R}_s^2r_s^1-
       \mathcal{R}_s^2r_s^2\right.\\
       && + r_s^1 \mathcal{R}_s^{1\dag} - r_s^1 \mathcal{R}_s^{2\dag} + r_s^1 \mathcal{R}_s^{2\dag}
       -r_s^2 \mathcal{R}_s^{2\dag}\\
       &&\left.\!+\mathcal{M}(r_s^1)-\mathcal{M}(r_s^2)\right|ds\\
       &\leq&
       \ds\int_0^t\left\{\left|\mathcal{R}_s^1-\mathcal{R}_s^2\right||r_s^1| + |r_s^1|
       \left|\mathcal{R}_s^{1\dag}-\mathcal{R}_s^{2\dag}\right|\right\}ds\\
       &&+\ds\int_0^t\left\{\left|\mathcal{R}_s^2\right||r_s^1-r_s^2|+|r_s^1-r_s^2|
       \left|\mathcal{R}_s^{2\dag}\right|\right\}ds\\
       &&+\ds\int_0^t\left\{\left|\mathcal{M}(r_s^1)-\mathcal{M}(r_s^2)\right|\right\}ds\\
       &\leq& C_1 \pa y^1-y^2 \pa_T +
       \ds\int_0^t C_2|r_s^1-r_s^2|ds
    \end{eqnarray*}
Here $C_1$ and $C_2$  depend on $\pa y^1\pa_T$, $\pa y^2\pa_T$ and
$T$. By Gronwall's lemma, we have
    \begin{eqnarray*}
    \vert r_t^1-r_t^2\vert &\leq& C_1 \pa y^1-y^2 \pa_T
    \exp\left\{\ds\int_0^t C_2 ds\right\}\\
    &\leq& C \, \pa y^1-y^2 \pa_T
    \end{eqnarray*}
for all $0 \leq t \leq T$, which implies \er{cts} as required.

\newpage 

\end{document}